\documentclass[reprint,superscriptaddress]{revtex4-1}
\usepackage{amsmath}
\usepackage{graphicx,dcolumn,bm,natbib,epsfig,verbatim,color,soul}
\usepackage{bm}        % for math
\usepackage{mathrsfs} % for scripted letters
\usepackage{color}
\usepackage{subfigure}  % use for side-by-side figures
\def\3dots{\:\raisebox{-0.5ex}{$\stackrel{\textstyle.}{:}$}\:}

\begin{document}
\title{Chase-and-run dynamics in cell motility and the
  molecular rupture of interacting active elastic dimers}
\author{David Mayett}
\affiliation{Department of Physics, Syracuse University, Syracuse, NY 13244, USA} 
\author{Nicholas Bitten}
\affiliation{School of Physics and Astronomy, Rochester Institute of Technology, Rochester, NY 14623, USA}
\author{Moumita Das}
\affiliation{School of Physics and Astronomy, Rochester Institute of Technology, Rochester, NY 14623, USA}
\author{J. M. Schwarz}
\affiliation{Department of Physics, Syracuse University, Syracuse, NY 13244, USA} 
\date{\today}

\begin{abstract}
 Cell migration in morphogenesis and cancer metastasis typically
 involves interplay between different cell types. We construct and
 study a minimal, one-dimensional model comprised of two different
 motile cells with each cell represented as an active elastic
 dimer. The interaction between the two cells via cadherins is modeled as a spring
that can rupture beyond a threshold force as it undergoes dynamic
loading via the attached motile cells.  We obtain a phase diagram
consisting of chase-and-run dynamics and clumping dynamics as a
function of the stiffness of the interaction spring and the threshold
force. We also find that while feedback between cadherins and
 cell-substrate interaction via integrins accentuates the chase-run behavior, feedback is not necessary
 for it. 
\end{abstract}
\maketitle
During embryonic development as well as in cancer metastasis, cells 
often undergo migration in groups~\cite{Friedl}. Such groups are typically composed
of cells of different types interacting with each other giving rise
to nontrivial migration modes.  For example, co-cultures of stromal
fibroblasts and carcinoma cells on top of an extracellular matrix (ECM)
reveal that the carcinoma cells move within tracks in the
ECM made by the fibroblasts~\cite{Gaggioli}.
 Another example of a
nontrivial migration mode occurs when neural crest (NC) cells and placodal (PL) cells are cultured next
to each other on a polyacrylamide substrate. The NC cells start chasing the PL cells via chemotaxis, while the PL cells run away from the NC cells when
contacted by them~\cite{Theveneau1}.  NC cells are highly multipotent cells that migrate extensively during embryogenesis, and eventually differentiate to
give rise to multiple cell types including some nerve and glial cells,
fibroblasts, and smooth muscle cells~\cite{Theveneau2}. Placodal cells
(PL), on the other hand, are embryonic cells that remain more
localized~\cite{Theveneau1}. They play a critical role in development of the cranial sensory system
in vertebrates\cite{Schlosser}.  

While there are a number of models of single cell migration or few
cell migration of the same cell type on
surfaces~\cite{Mogilner, Keren, Barnhart, Kumar, Loosely,Tjhung,Camley,Sergerer,Sun}, the rules
governing the interplay between different cell types from a
cell migration standpoint remain largely unknown. Inspired by the
NC/PL cell experiment~\cite{Theveneau1}, we consider a minimal, one-dimensional
model of two different, but interacting, cells.  Each cell is modeled as an active elastic dimer with
focal adhesions acting as catch bonds at the leading edge of a
crawling cell
and slip bonds at its rear~\cite{Lopez}.  The mechanosensitive activity, which is
incorporated as a changing equilibrium spring length depending on the
loading state of myosin, combined with the catch/slip bond
asymmetry generates motion even in the absence of broad lamellipodia 
typically observed in cells crawling along two-dimensional surfaces~\cite{Keren}.  By invoking a minimal set of assumptions for the
interaction between the two cells in our one-dimensional model, we can predict, in principle, all 
possible migration outcomes and, therefore, begin to classify the rules of interplay
between two motile cells. More
specifically, we can observe nontrivial migration modes
such as the chase-and run phenomenon and ultimately distinguish
between various mechanisms for contact-inhibition-locomotion
(CIL)---motion in which two cells move towards each other, collide,
and then move away from each other~\cite{Theveneau1,Abercrombie, Carmona}.  Both behaviors are fundamentally
one-dimensional and can therefore be captured with our 
one-dimensional model capturing mesenchymal migration along a taut ECM
fiber, for example.

\begin{figure}[th]
\centering
\includegraphics[width=0.9\linewidth]{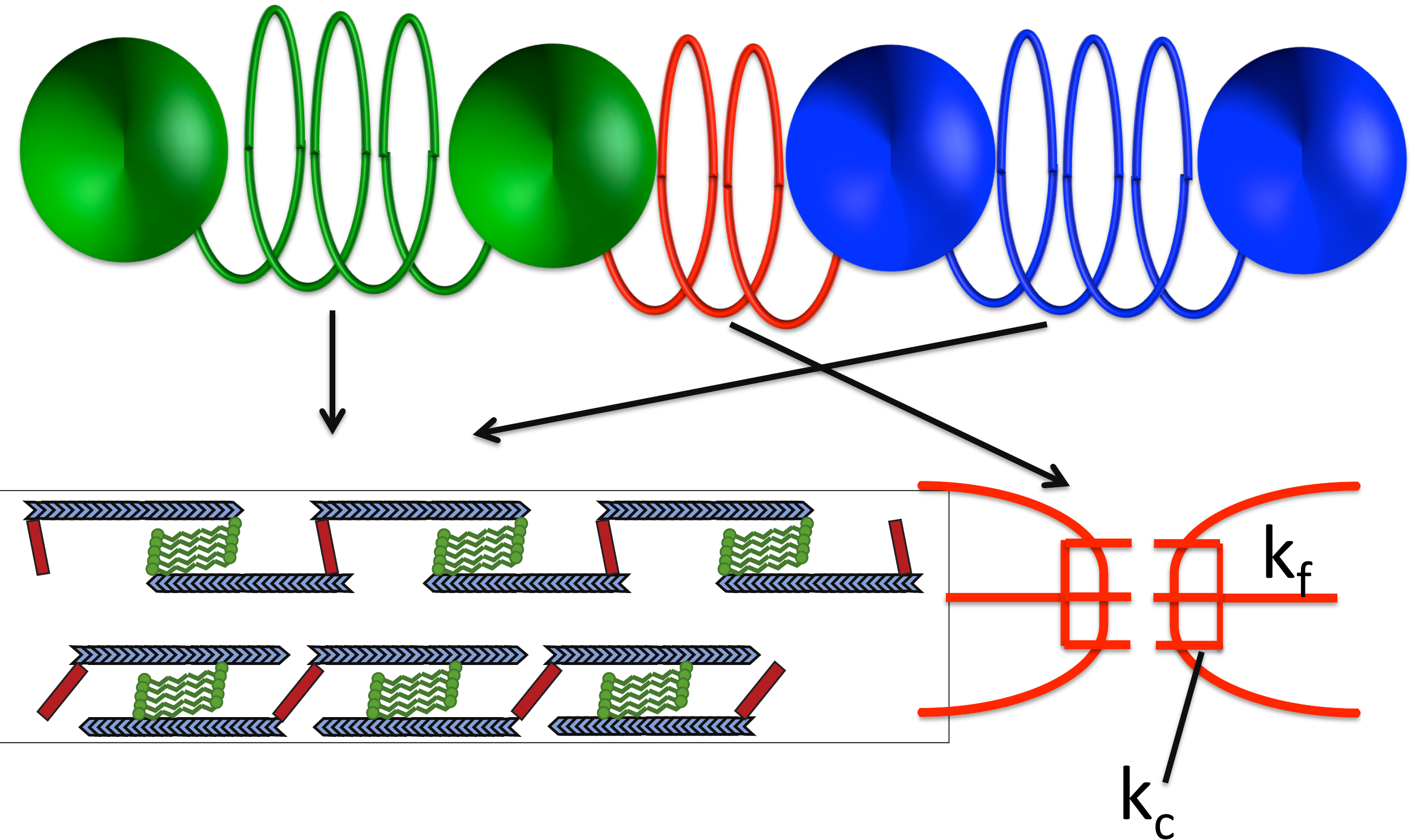}
\caption{\label{fig:dimers}(Color Online) Schematic representation of
  two cells (blue and green) with the red interaction spring containing $N$ cadherin
  molecules in parallel, each with spring stiffness $k_c$ and a
  filopod with spring stiffness $k_f$. Each blue and green spring
  represents the prominent stress fibers along the length of the
  cell and, therefore, is active with both the extended mode (top) and
  contracted mode (bottom). The blue filaments represent actin filaments, red rectangles, alpha-actinin, and the green shapes, myosin minifilaments.   }
\end{figure}

{\it Model:}  Cells moving along ECM fibers or patterned microchannels 
extend themselves along the fiber/channel and move even in the absence of
lamellipodia~\cite{Friedl2,Yamada,Kumar,King,Fraley}.   Pronounced stress fibers are a
characteristic feature of this migration mode called mesenchymal migration~\cite{Friedl2}. Stress
fibers primarily consist of actin filaments, myosin, and a passive cross-linker alpha-actinin~\cite{Fiber}.  
Structurally, they can be though of as made of parallel arrangements of acto-myosin units in series, where
each actomyosin unit may be considered as two actin filaments connected by a myosin mini-filament, and the crosslinker
alpha-actinin at each end (Fig.~\ref{fig:dimers}). 
We have previously developed a model to describe a single cell in this migration mode~\cite{Lopez} which
serves as the foundation for our investigation of two co-migrating interacting cells. 

The main ingredients of the single cell model, which has been studied
in detail in ~\cite{Lopez} and reviewed in the SI, are as follows:\\
(1) The migrating cell is modeled as two beads connected by an active
spring. The spring represents the stress fiber and the beads denote the
 location of focal adhesions at positions, say, $x_1$ and $x_2$, which enable the stress fibers connect
 to the ECM fiber.\\
(2) The active spring has two different equilibrium lengths,  $x_{eq1}$ and $x_{eq1}-x_{eq2}$, corresponding to the nearly unloaded and loaded states of myosin. The former is determined mainly by passive alpha-actinin driven extension and the latter by active myosin driven contraction.   
The equilibrium spring length, $x_{eq}$, can then be written as 
$x_{eq}=x_{eq1}-x_{eq2}\Theta(x_1-x_2-l)$ where $\Theta(x_1-x_2-l)$ is
the Heaviside step function. The transition between the two modes is
determined by the extension of the spring: the larger the extension, 
the more the tensile load on myosin thereby inducing contractility of myosin given its catch bond
nature~\cite{Guilford}. \\
(3) There exists hysteresis in the equilibrium spring length with
$l=l^{\uparrow}$ as the active spring extends and
$l=l^{\downarrow}\ne l^{\uparrow}$ as it compresses due to
increasing overlap between actin filaments allowing for more passive
crosslinking by alpha-actinin, potential conformational changes in
the alpha-actinin, and internal frictional losses. \\
(4) Integrins, one of the principal proteins in focal adhesions
adhering the cell to the fiber~\cite{Waterman}, can act as catch bonds under
 repeated loading~\cite{Kong}.  They are more likely to act as catch bonds
  at the leading edge of a crawling cell due to the more dynamic environment for the maturation of focal
 adhesions, while at the rear they act as typical
 slip bonds where focal adhesions are merely being
 disassembled. So, at the front of the cell, the initiation of
 focal adhesions call for a ``small'' friction coefficient, but once
 the focal adhesions form and develop, the friction increases. This ``catching''
 mechanism of cell-track adhesion allows the cell's front to expand
 and explore new territory and after having done that, then allows for
 the cell's rear to retract with the cell front not losing grip on the
 new territory it just explored.  So
 we define the friction coefficient at the leading edge to be   
$\gamma_1=\gamma_{11}+\gamma_{12}\Theta(x_1-x_2-l^{\uparrow(\downarrow)})$
with $\gamma_{11},\gamma_{12}>0$ and $\gamma_{11}<\gamma_{12}$. Because the integrins track
myosin activity, the hysteresis exhibited by myosin is also
exhibited in the friction. Finally, $\gamma_{2}$, the
friction coefficient for the now ``rear'' bead, is assumed to be
constant with the integrins acting as ordinary slip bonds.  \\
(5) The combination of activity that depends on the strain in
the stress fiber and the asymmetry of the focal adhesions
at the leading and rear edges leads to directed cell motion in the
direction of larger friction~\cite{Dembo}.

To address the interaction between two motile cells in one-dimension,
each cell is described by above single cell model. To be concrete, the beads are described by their positions  $x_{i}(t)$, with $i\in
[1,4]$, where $i=1$ denotes the rightmost bead and $i=4$ the
leftmost. The focal adhesions associated with the $i^{th}$ bead are
denoted by $\gamma_{i}$. For the cell on the right, $\gamma_{1} =
\gamma_{2} $. This cell is stationary given the symmetry in the
friction, provided no outside forces act on it. This is our model PLL (placode-like)
cell.  As for our neural crest-like (NCL) cell (cell on the left), the action of chemotaxis is
implicitly described by the breaking of the symmetry between the rear
and front bead focal adhesion of the left cell to generate directed
motion. Thus, for the cell on the left we have $\gamma_{3} =
\gamma_{33}+\gamma_{34}\Theta(x_{3}-x_{4}-l^{\uparrow(\downarrow)})$
and $\gamma_{4}$ is a constant. Both cells have changing equilibrium
spring lengths denoted by $x_{eq} =
x_{eq1}-x_{eq2}\Theta(x_{1}-x_{2}-l^{\uparrow(\downarrow)})$ (for the
PLL cell) to incorporate myosin driven contractility and $\alpha-$actinin driven extensibility as described in~\cite{Lopez}.

The cell-cell interactions are mediated by cadherin molecules.  These
molecules localize at the ends of filopodia (small actin-bundle-based
protrusions) demonstrating that cadherins also interact with
the actin cytoskeleton~\cite{Yap}. The number of cadherin molecules
at the tips of filopodia and other actin-based protrusions range from
hundreds to thousands.  We assume that cadherin molecules, each
modeled as a linear spring with spring constant $k_c$, bind in
parallel and are then bound to a filopod also modeled as another
linear spring with spring constant $k_f$. See Fig. 1. When the two cells come in close enough proximity,
an interaction spring forms between them. This proximity is denoted
by $l_a$. Because the two cells have their own inherent dynamics, they
can in principle pull on the cadherin bonds and rupture them \cite{Seifert}. For
simplicity, we assume the interaction spring
can rupture when $k_f(x_2-x_3-l_{eq})>N_0f_c$, where $f_c$ is the
critical force threshold that will rupture an individual cadherin
bound for $k_f\approx k_c$ with $k_f=k$ for notational ease. Rupture can only occur when the two
beads at either end of the interaction spring are moving away from
each other.

\begin{figure*}[htp]
\centering
\subfigure{\includegraphics[scale=0.235]{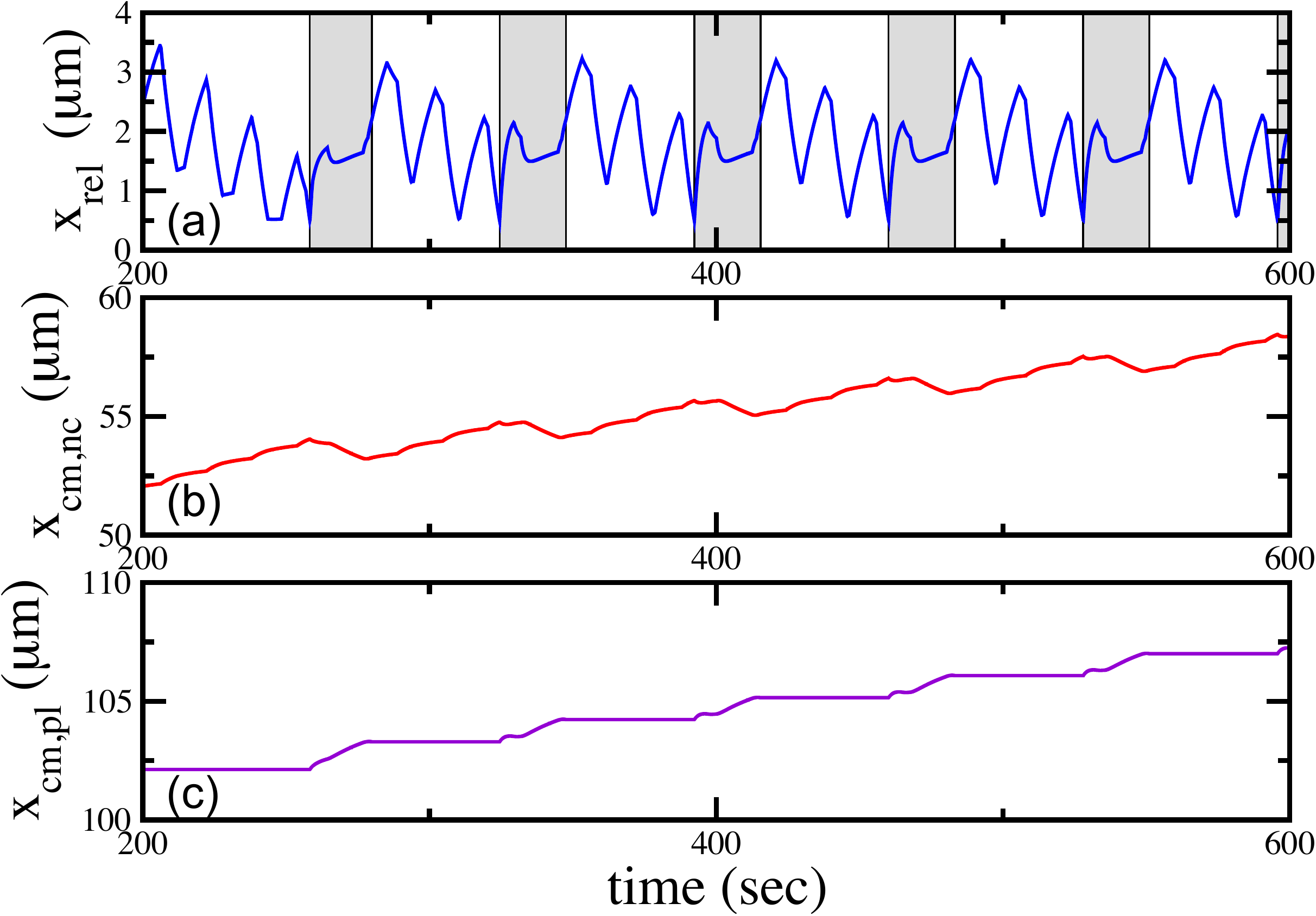}}
\subfigure{\includegraphics[scale=0.235]{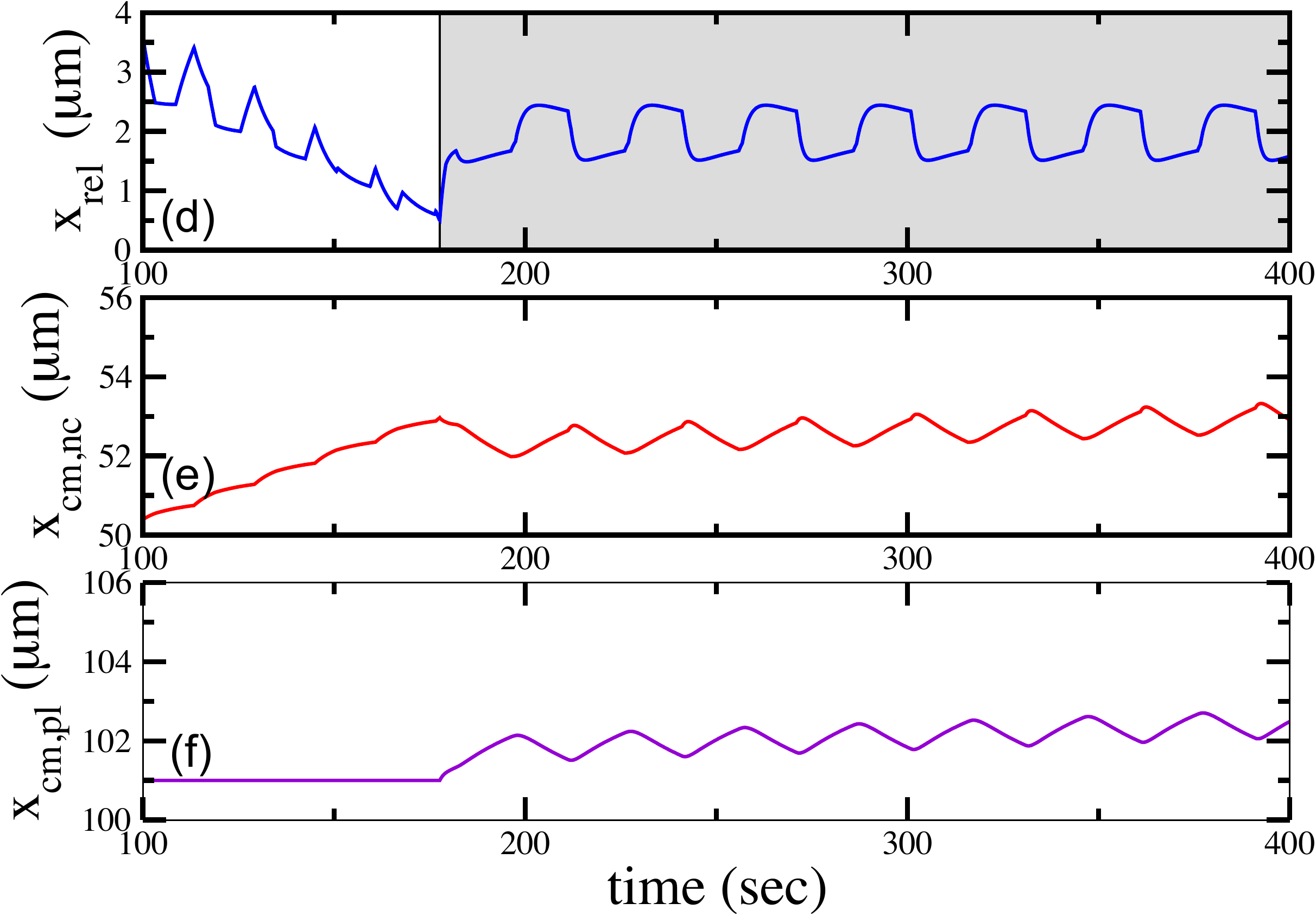}}
\subfigure{\includegraphics[scale=0.235]{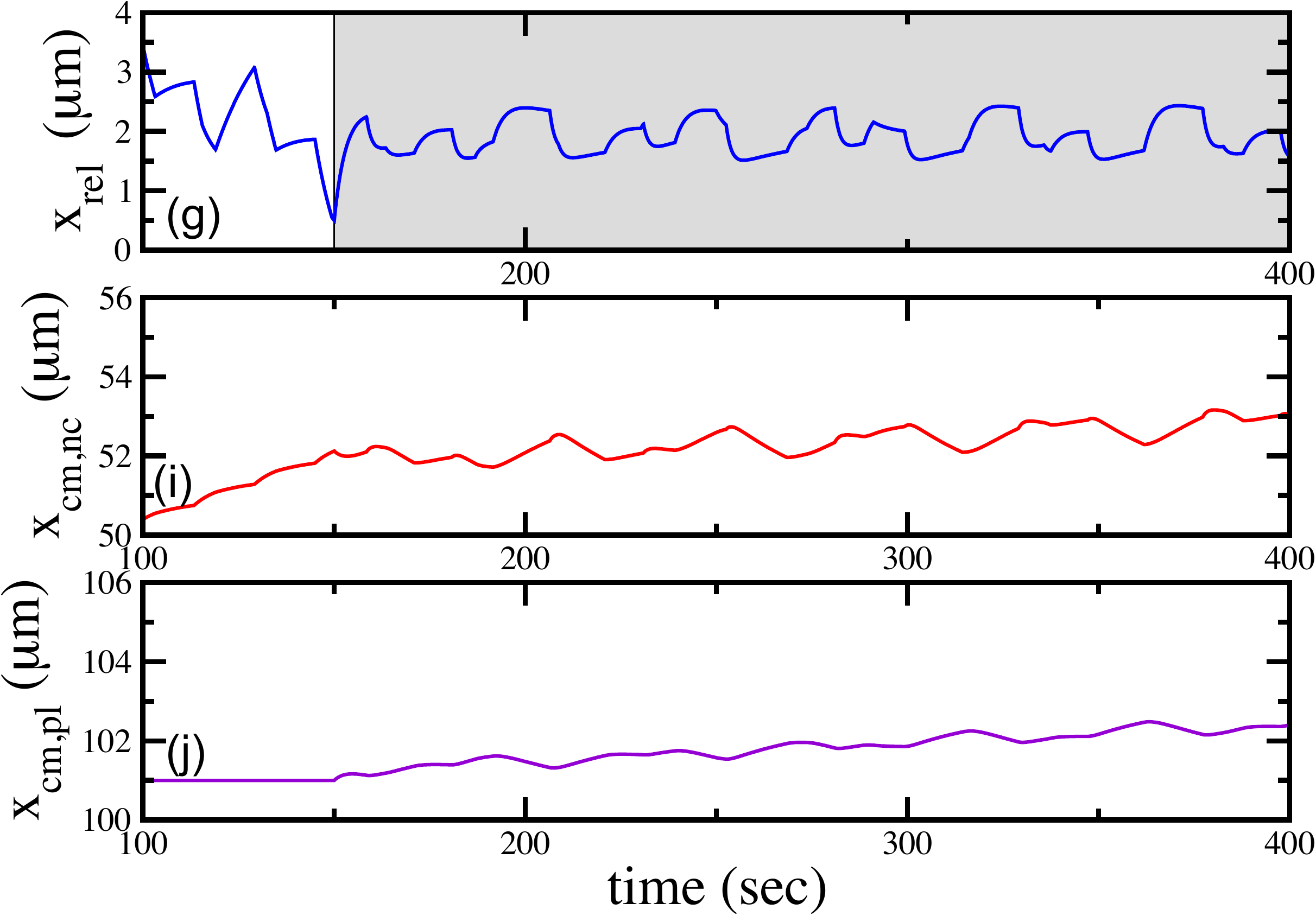}}
\caption{\label{fig:set1} (Color Online) (a)-(c) The relative distance between the two
  cells, the center of mass position of the neural crest, and the
  center of mass position of the placode cell, all as a function of
  time.  Here, $f_r=0.01\, nN$ and $k=5\, nN/\mu m$.  The grey region
  in the top figure indicates when the interaction is tuned
  on. (d)-(f) The
  same as (a)-(c) but with $f_r=0.03\, nN$ and $k=5\, nN/\mu m$. (g)-(i) Here,
  $f_r=0.03\, nN$ and $k=5\, nN/\mu m$ but with larger friction
  coefficients for the PLL cell. }
\end{figure*}

Putting together the different components of the model, the four coupled equations of motion of the beads are as follows:
\small
\begin{eqnarray}
     \begin{aligned}
      & \gamma_{1}(x_{1},x_{2},l^{\uparrow}, l^{\downarrow})\dot{x}_{1}(t) =\nonumber- k_{1}\left[x_{1}-x_{2}-x_{eq}(x_{1},x_{2},l^{\uparrow},l^{\downarrow})\right]\nonumber\\&+ \sqrt{A_{1}}\zeta_{1}(t) \\
       &\gamma_{2}(x_{1},x_{2},l^{\uparrow}, l^{\downarrow})\dot{x}_{2}(t) =\nonumber k_{1}\left[x_{1}-x_{2}-x_{eq}(x_{1},x_{2},l^{\uparrow},l^{\downarrow})\right]\nonumber\\& -k_{c}[x_{2}-x_{3}-l_{eq}]+ \sqrt{A_{2}}\zeta_{2}(t)\\
       &\gamma_{3}(x_{3},x_{4},l^{\uparrow}, l^{\downarrow})\dot{x}_{3}(t) =\nonumber - k_{2}\left[x_{3}-x_{4}-x_{eq}(x_{3},x_{4},l^{\uparrow},l^{\downarrow})\right]\nonumber\\& +k_{c}[x_{2}-x_{3}-l_{eq}]+ \sqrt{A_{3}}\zeta_{3}(t) \\
       &\gamma_{4}(x_{3},x_{4},l^{\uparrow}, l^{\downarrow})\dot{x}_{4}(t) = k_{2}\left[x_{3}-x_{4}-x_{eq}(x_{3},x_{4},l^{\uparrow},l^{\downarrow})\right]\nonumber\\& +\sqrt{A_{4}}\zeta_{4}(t).
     \end{aligned}
\end{eqnarray}
\normalsize
For completeness, we have included fluctuations denoted by
$\sqrt{A_i}\zeta_i(t)$, where $\zeta_i(t)$ is a Gaussian random
variable with $<\zeta_i(t)>=0$ and
$<\zeta_i(t)\zeta_j(t')>=\delta_{ij}\delta(t-t')$. These fluctuations
are due to activity and are not related to any temperature via a
fluctuation-dissipation theorem. We will ultimately study the limit $A_1=A_2=A_3=A_4=A$. We have independent estimates for all
but three parameters based either on experiments or prior modeling
discussed in Ref.~\cite{Lopez} or elsewhere. Specifically,
$k_1=k_2=1 \,nN/\mu m$, $x_{eq1}=50\,\mu m$,  $x_{eq2}=5\,\mu m$, $l^{\uparrow}=48.5 \,\mu m$,
    $l^{\downarrow}=46.5 \,\mu m$, $\gamma_{11}=20 \,nN s /\mu m$, 
    $\gamma_{12}=0\,nN s /\mu m$, 
 $\gamma_{33}= 10 \,nN s /\mu m$, 
    $\gamma_{34}=20 \,nN s /\mu m$, 
    $\gamma_{2}=20\,nN s\, /\mu m$, 
    and $\gamma_{4}=20\,nN s /\mu m$.   
For the interaction parameters, we know from single molecule
experiments that $f_c=40\,\,pN$ for N-cadherin and $f_c=70\,\,pN$ for
E-cadherin~\cite{Panorchan}, $k=k_f$ is of order $1\,nN/\mu
m$~\cite{Bornschlogl} (since $k_c\approx k_f)$,
and $N_0$ is of order 100 per pseudopod~\cite{Quang}. The only
parameters we do not have independent estimates for are $l_{eq}$, $l_a$, and
$A$, though $l_{eq}$ and $l_a$ are determined by the appropriate
lengthscales in the system.  We set $l_a=0.5\,\,\mu m$ and vary both $l_{eq}$ and $A$.  

To study this model, we implement 4th order Runge-Kutta integration
scheme in the absence
of noise.  With noise, we implement a Euler-Marayuma integration
scheme. We have checked our simulations against the analytical solution
for some parameter values.  For the single active dimer, there were
two analytical solutions to be pieced together according to the cell's 
history.  For the interacting active elastic
dimer case there are eight analytical solutions to be pieced
together according to each cell's history---two for each cell and two more cases for the interaction spring
either on or off.  Given the plurality of solutions, the majority of
our results are based on simulations.

\begin{figure}[b]
\centering
\includegraphics[width=0.8\linewidth]{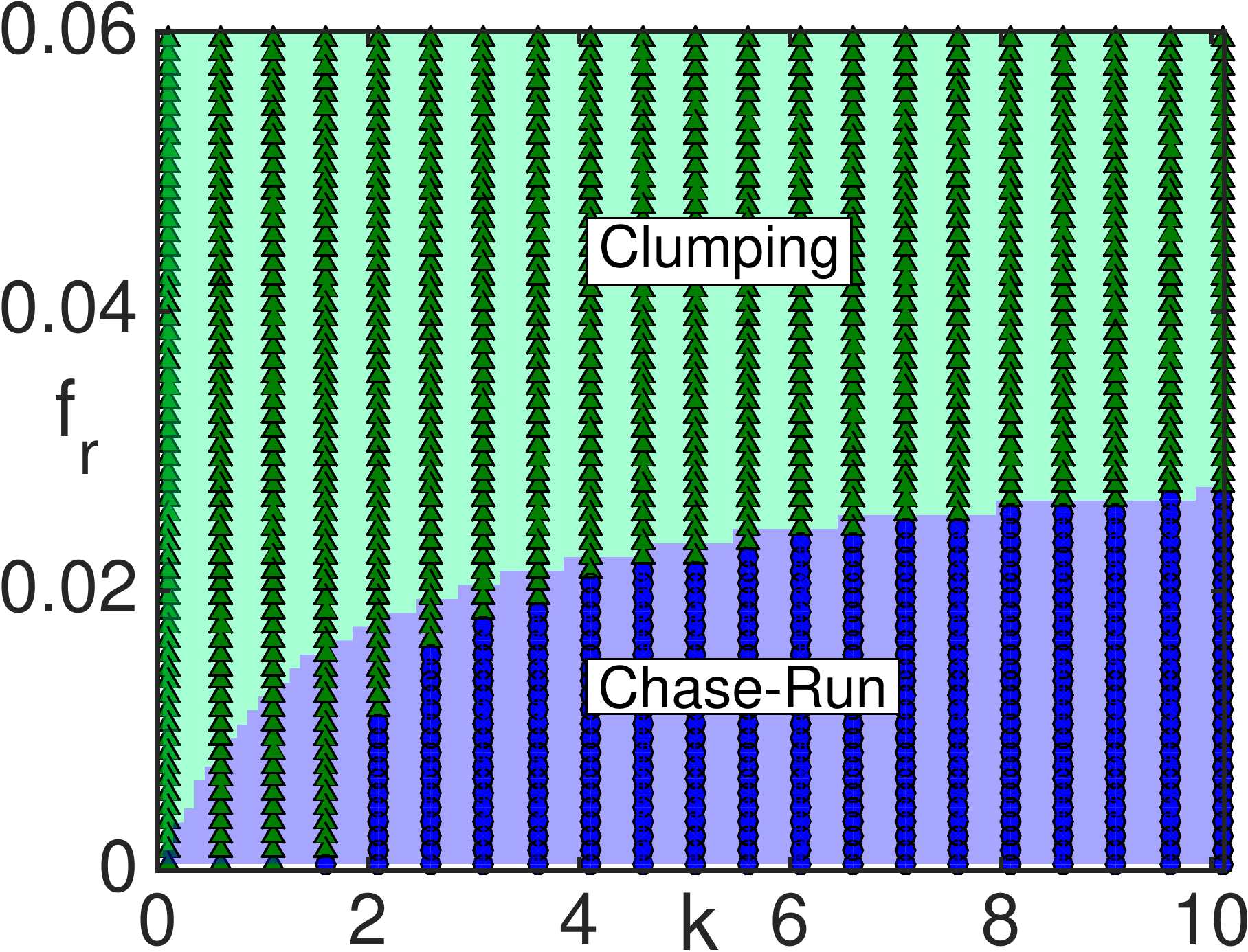}
\caption{\label{fig:phasedia1} (Color Online) Chase-Run and Clumping states for the two-cell model
 for the parameter values noted in the text. The symbols, blue circles
 (Chase-Run) and green triangles (Clumping) indicate simulation data,
 while the corresponding blue and green shaded regions correspond to
 the analytical result.  The units of $k$ are 
$nN/\mu m$ and the units of $f_r$ are $nN$.}
\end{figure}

\begin{figure*}[htp]
\centering
\subfigure{\includegraphics[scale=0.23]{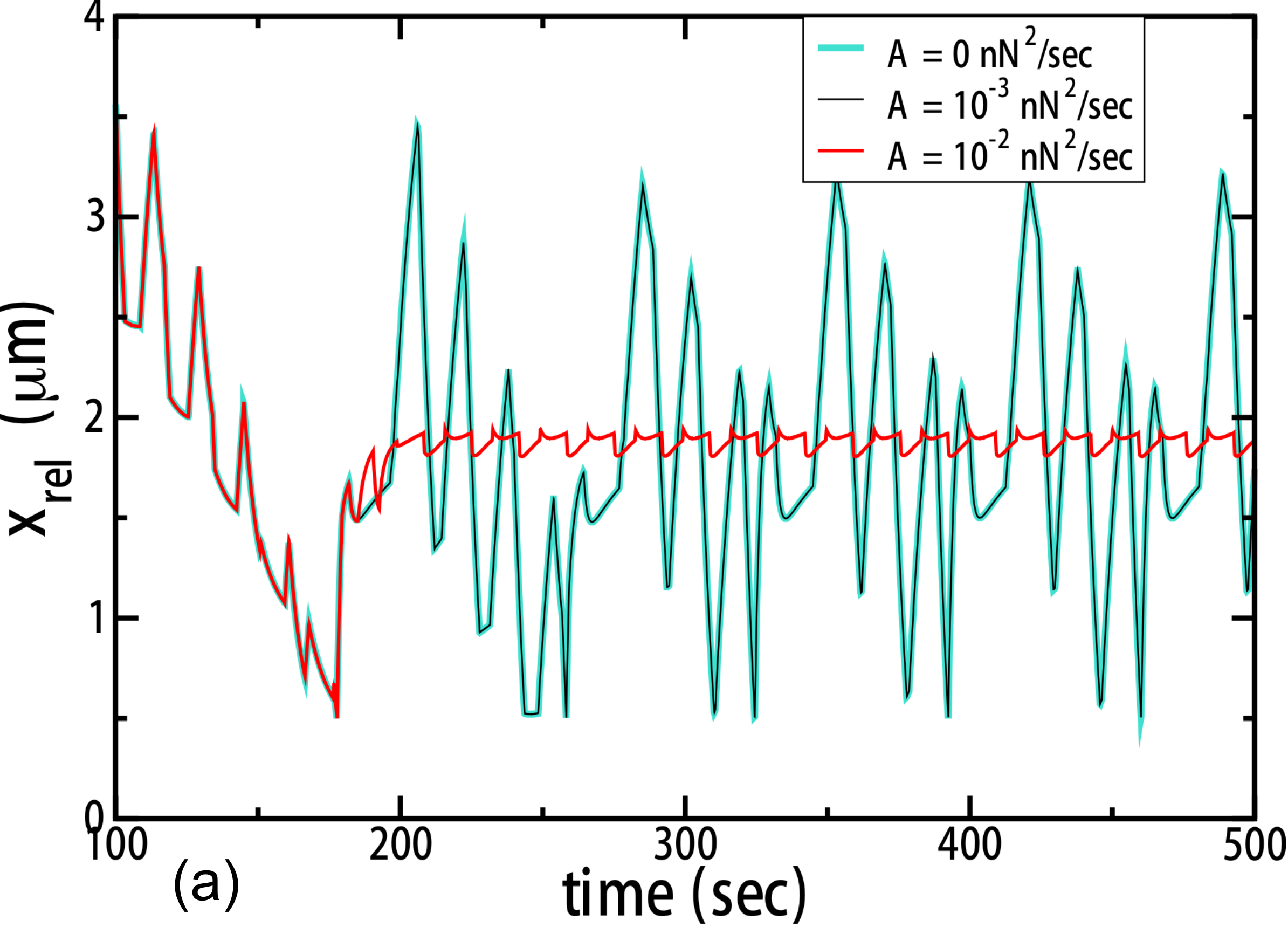}}
\subfigure{\includegraphics[scale=0.23]{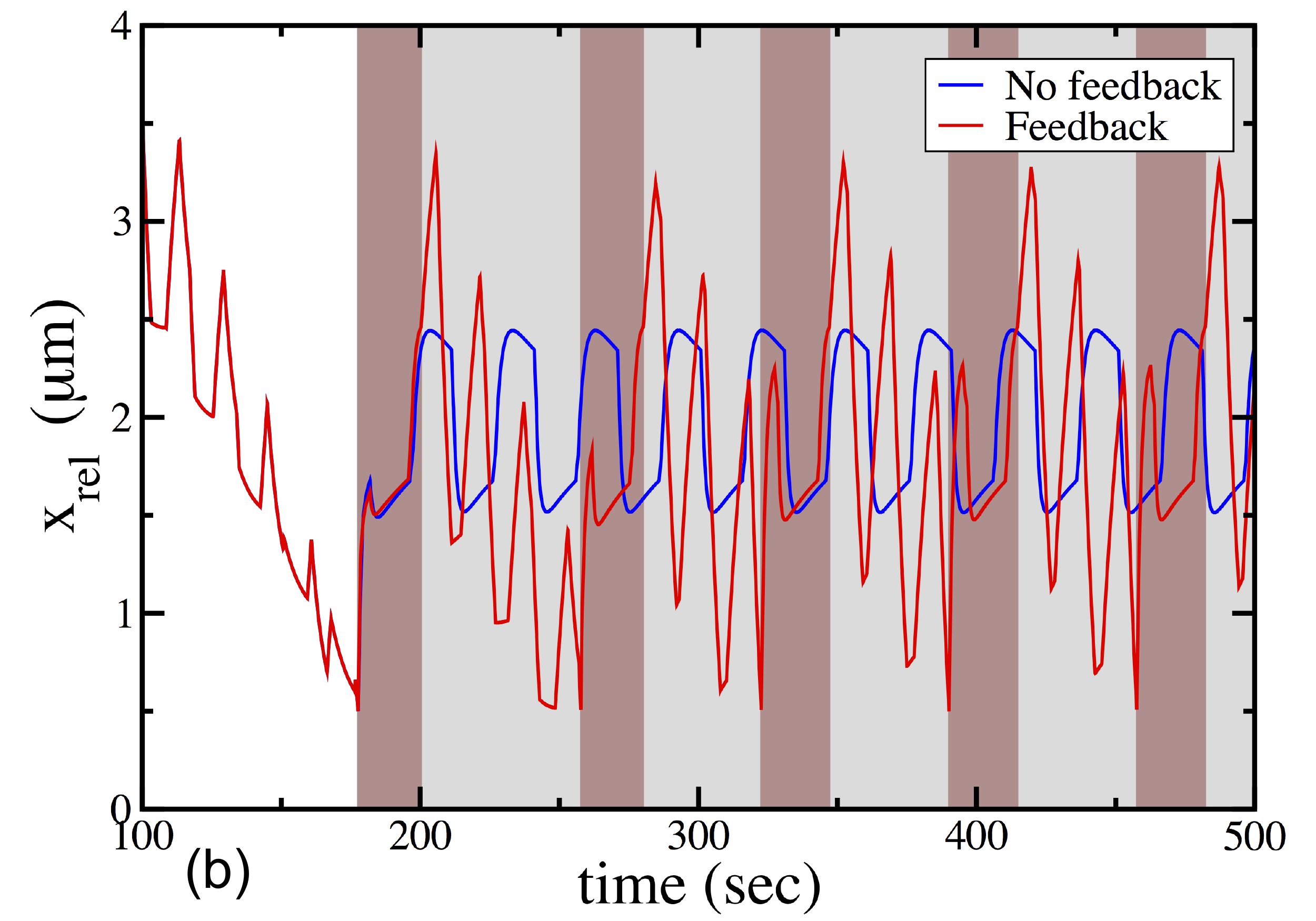}}
\subfigure{\includegraphics[scale=0.23]{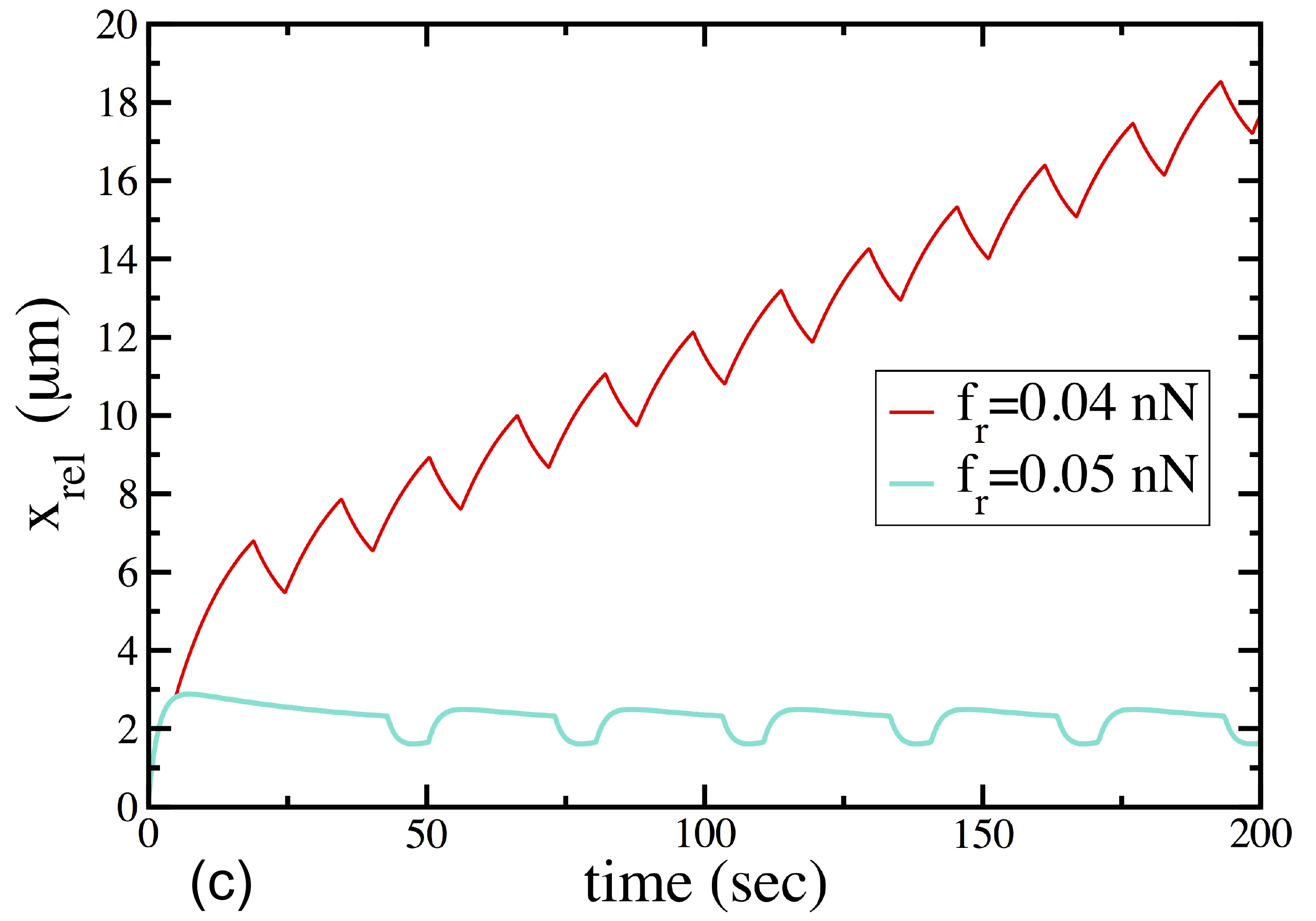}}
\caption{\label{fig:feedback}(Color Online) (a) The relative distance between the cells
  with and without noise with variance $A$ on each of the four
  beads  (b) The relative distance between the two
  cells with and without feedback.  The brown shading represents the
  presence of the interaction spring in the chase-and-run case, the
  gray, the clumping case. (c) Two cells moving apart from each other are not always able to
 rupture the interaction spring, i.e. escape. It depends on the
  rupture force.}

\end{figure*}

{\it Results:} To classify the types of interactions between the two
different cells, we study the cell dynamics as a function of the
junction spring stiffness, $k$, and the rupture force between cadherin
molecules.  We focus on $x_{rel}=x_2(t)-x_3(t)$,
$x_{cm,pl}(t)=\frac{1}{2}(x_1(t)+x_2(t))$, and
$x_{cm,nc}(t)=\frac{1}{2}(x_3(t)+x_4(t))$. We initialize the NCL cell some distance away from the PLL cell and iterate until
they interact. As a result of the asymmetry in the friction coefficients of the NCL cell, it will
migrate towards the PLL cell, mimicking the movement of  the NCL cell toward the PLL
cell due to chemotaxis, or a chemical gradient. The PLL cell, on the
other hand, does not move (on its own) since there is no asymmetry in
its friction coefficients. Figure 2(a)-(c) plots these quantities for
$f_r=0.01\, nN$ and $k=5\, nN/\mu m$ as they interact. For these
particular values, the cell springs are able to rupture the
interaction spring, i.e. separate. But as the NCL cell, again, moves
toward the PLL cell, the two cells interact again and the process
repeats ad infinitum. We classify this dynamic state as {\it
  chase-and-run} behavior since the interaction spring is ruptured
with the PLL cell pulling away from the NCL cell. Note that the
position center of mass of the PLL cell only changes when in contact
with the NCL cell.  

Now we increase the rupture force to $f_r$ to $0.03\, nN$. See
Fig. 2(d)-(f). At this increased rupture force for the cadherin
molecules, the interaction spring always remains on, i.e. the two
cells never separate once they interact.  We dub this dynamic state as
{\it clumping}.  In the presence of the chemotaxis and one sedentary
cell, chase-and-run and clumping are the two behaviors one can observe in  terms of how the cells come into contact. 
If we increase the two friction coefficients of the PLL cell such that the time scales are different for each cell, for $f_r=0.01\, nN$ and
$k=5\, nN/\mu m$, we observe quasiperiodic behavior in the relative
distance between the cells. See Fig. 2(g)-(i).

To summarize our findings in terms of searching for chase-and-run and
clumping dynamics as a function of the interaction spring stiffness
and the rupture force, we present a phase diagram as a function of $k$
and $f_r$ in Fig.~\ref{fig:phasedia1}. The system transitions from chase-and-run at smaller
rupture force to clumping at larger rupture forces.  As the
interaction spring stiffness increases far beyond the cell spring
stiffness, the energetics is dominated by the interaction spring and
the dependence on the rupture force on the transition decreases. We
can estimate the transition line by looking at the case where each
cell spring is in its contracting phase (smaller equilibrium spring
length) so that each cell spring maximally pulls on the interaction spring to
potentially rupture it. 

We also investigate the model in the presence of active noise due to the presence of
fluctuations in the myosin motors, for example. See
Fig.~\ref{fig:feedback}(a). We find that the phase-diagram in
Fig.~\ref{fig:phasedia1} is robust to small fluctuations (see
Fig.~\ref{fig:feedback}(a) for an example). However, a
system undergoing chase-and-run dynamics in the absence of noise can
be driven to clumping with large enough fluctuations. 
We have assumed here uncorrelated, or Gaussian noise, for simplicity.
Should active noise be an important contribution, we anticipate
fluctuations that correlate with motor activity, so that correlated
noise may indeed be a more accurate representation of the
biomechanics. 

We now discuss the phase diagram in the context of the NC-PL
experiments~\cite{Theveneau1}. The authors of Ref.~\cite{Theveneau1}
conjectured that the switching
of $N$- to $E$-cadherin binding drove the system from chase-and-run to
clumping dynamics.  We observe that here as well
within the appropriate force scale.  As mentioned earlier, the rupture
force for N-cadherin is approximately $40 \, pN$, while for
E-cadherin, it is approximately $70\, pN$.  We observe, for example
for $k=2$, the doubling of rupture from $10 \,pN$ to $20 \, pN$ drives
the system from chase-and-run to clumping.  The experimentalists also
conjectured that feedback between the cadherin and integrin is 
important for the chase-and-run dynamics -- the more cadherin bind, the less
integrin bind~\cite{Theveneau1}. We, however, observe chase-and-run behavior 
even without any feedback between the two types of molecules.  We can, of
course, incorporate this feedback into our model as follows. 
If the interaction spring is on, the friction coefficients on both
sides of the spring are decreased, say, by half (in both states for the
NC cell).  With this feedback, we observe that the chase-and-run state 
occupies a larger part of parameter space.  For instance, with no feedback,
the transition for $k=5\, nM/\mu m$ occurs at $f_r=0.021 \, nN$ but
with the feedback, the transition occurs at $f_r=0.023\,
nN$. Alternatively, a clumped system with no feedback can be driven
to the chase-and-run state with feedback. See Fig.~\ref{fig:feedback}(b). 

Finally, we address the issue of polarity.  Polarity, here, is
determined by the asymmetry in friction. If two cells initially moving toward
each other, interact and then change polarity, the relative distance
between them would decrease as they meet, and then increase as
they interact and reverse direction.  This behavior is known as
contact-inhibition-locomotion (CIL)~\cite{CIL}. We conjecture that 
feedback between the cadherin and integrin binding could drive
the cell to change its polarity and, therefore, potentially reverse
direction. If the integrin binding becomes weaker one
side of the cell due to molecules participating more in the cadherin
junctions than in the focal adhesions, then integrin binding on the other side of the
cell may increase to compensate.  This increase in ultimately friction
on the other side of the cell may be enough to begin to generate
motion away from the ``other'' cell. If the two cells rupture the
interaction spring between them, the two cells each go  ``their merry
way''. Therefore rupture is an important part of the process. Within
our model, it turns out that
cells cannot always rupture the interaction spring between them, even
if both cells are moving away from each other.  See Fig. 4(c).  This
is counterintuitive at first but makes sense since the interaction
spring does not allow the cell springs to transition as readily between the two contracting and extending states so that each cell cannot escape each other.

{\it Discussion:} We have developed a one-dimensional mescoscopic model to describe the interaction between 
two cells mediated by N/E-cadherin. Like the experiments~\cite{Theveneau1}, we observe a transition from
chase-and-run dynamics to clumping dynamics when switching from N- to
E-cadherin. In the chase-and-run case, the NC cell acts as elastic
herder controlling the motion of the PLL cell by interacting with
it. This herding sheep analogy is distinct from the
horse-carrot analogy presented in Ref.~\cite{Theveneau1}. 
With our phase diagram, we can predict which behavior
will occur depending on the rupture force of a single binding molecule
that can be tested with genetic modification of both types of
cadherins.  We have also addressed the types of interactions two different
motile cells can have in one-dimensional migration.  Our model can be adapted
to groups of NCL and PPL cells with each cell described as a group of
active springs and there being interaction springs between
each cell with the interaction springs between the NCL and PLL having
a lower rupture threshold than the interaction springs between two PLL
cells. 

Our model connects molecular and cellular scales to provide a mechanistic understanding of collective migration of heterogeneous cell populations that combine mesenchymal migratory properties and cadherin based cell-cell junctions. It may, therefore, not only apply to the enhanced migration of neural crest cells during morphogenesis, 
but also provide insights into the microscopic mechanical interactions between co-migrating cancer cells and non-tumorigenic cells, 
which are known to have significantly different mechanical and adhesion properties~\cite{co-culture}. 
Finally, our results demonstrate that a quantitative framework of cell-cell interaction should include molecular rupture forces~\cite{Seifert} as well as the mechanosensitive activity
of the cytoskeletal machinery to help inform the case of more than two interacting cells with varying degrees of cell motility, thereby quantifying the coordinated migration of cells. 

DM and JMS acknowledge support from DMR-CMMT-1507938. NB and MD were partially supported by a Cottrell College Science Award from Research Corporation for Science Advancement.
%\bibliography{twocellbibliography}

\end{document}